\documentclass[twoside]{article}
\usepackage[latin9]{inputenc}
\usepackage[letterpaper]{geometry}
\usepackage{color}
\usepackage{array}
\usepackage{multirow}
\usepackage{amsmath}
\usepackage{graphicx}
\usepackage{esint}

\makeatletter

\providecommand{\tabularnewline}{\\}

\usepackage[scanall]{psfrag}
\usepackage{color}

\makeatother

\begin{document}

\title{\textcolor{black}{Using a 1D bandgap metamaterial to build up a delay
line.}}

\author{\textcolor{black}{J Lucas$^{*}$, E Geron$^{*}$, T Ditchi$^{\dagger}$}\\
\textcolor{black}{Laboratoire de Physique et d'étude de la matière.}\\
\textcolor{black}{{*} PSL Research University, ESPCI-ParisTech}\\
\textcolor{black}{$\dagger$ Sorbonne Universités, UPMC Univ Paris
06}\\
\textcolor{black}{CNRS UMR8213 }\\
\textcolor{black}{F-75005 Paris, France}\\
}

\maketitle
\textcolor{black}{\medskip{}
}

This paper is a postprint of a paper submitted to and accepted for
publication in IET Microwaves, Antennas \& Propagation and is subject
to Institution of Engineering and Technology Copyright. The copy of
record is available at IET Digital Library.

\section*{\textcolor{black}{Abstract}}

\textcolor{black}{Classical lumped components delay lines are built
with inductances and capacitors arranged as to match the standard
telegraphist line model to obtain slow group velocity lines. In this
paper we demonstrate that a practical way to obtain significant group
velocity reduction consist in taking advantage of the forbidden bands
that present a 1D meta-material. The feasibility of this kind of line,
as well as its ability to transmit data are discussed and practically
verified.}

\section{\textcolor{black}{Introduction}}

\textcolor{black}{The great historical application of delay lines
is the radar application. They are used to separate the echoes of
moving targets from the echoes of static objects such as the ground
itself. Actually, long delays are obtained in radar applications by
converting the radar frequency signal to optical signal which allows
very long delays by making it propagate in wound optical fiber before
converting it back to microwaves frequencies for processing. Another
solution would consist in using shorter media where the velocity is
much lower. Without dispersion, the only way to obtain a slow velocity
in a line or a material consists in using high dielectric constant
material. Unfortunately, without changing the nature of the waves
as with SAW delay lines \cite{Schultz1970}, only small velocity reduction
can be obtained this way. It is possible though to obtain large velocity
reduction by taking advantage of dispersion. Many applications are
not wide band and are therefore compatible with smaller bandwidth.
It is possible to obtain a significant velocity reduction in a medium
over a limited bandwidth by taking advantage of some resonance. Many
works based on this effect have been carried out such as in \cite{Lucyszyn1995},
or more recently in optics~\cite{Junker2007,Toshihiko2008,Thevenaz2008,Dowlett2013}.
In this paper, we propose to use the properties of bandgap metamaterial
lines~\cite{Trytiakov2007} to obtain the same result focusing on
a propagation approach.}

\textcolor{black}{Firstly, we discuss the interest of relatively narrow
bandwidth low group velocity line for delay line applications. We
demonstrate how to calculate the required gain to obtain a given group
velocity reduction using the Kramers-Kronig relations~\cite{Kronig1926,Bohren2010,Lucas2012}
and how it applies to 1D metamaterials. Then, we expose the theoretical
key features that control the behavior of such lines : characteristic
impedance, group velocity and bandwidth, and how they interact with
each others. The design of the line is then tuned using a circuit
simulation software~\cite{ADS2012} that allows to take into account
the influence of the actual components implementation. At this point,
the losses in both substrate and lumped components as well as impedance
matching with standard 50~$\Omega$ lines are taken into account.
Finally the results obtained with a prototype are presented. The usability
of such a system to obtain large delays is verified and discussed
focusing on ability to transmit data. }

\section{\textcolor{black}{Delay lines\label{sec:Delay-lines_General}}}

\textcolor{black}{An ideal electrical delaying device is a device
whose frequency response is : ${\displaystyle H(\omega)=e^{-j\omega t_{0}}}$.
In this equation, $t_{0}$ is the delay, $\omega=2\pi f$ the circular
frequency, and $j=\sqrt{-1}$. In the case of a line, the delay $t_{0}$
is due to the propagation in the line and the frequency response can
be rewritten as: $H(\omega)=e^{-j\frac{\omega}{v}L}$, where $L$
is the dimension of the line and $v$ a celerity which can be written
as : $v=\frac{L}{t_{0}}$. As $t_{0}$ and $L$ are constants, $v$
is a constant as well. This corresponds to a non dispersive medium.}

\textcolor{black}{}
\begin{figure}[h]
\centering{}\textcolor{black}{\fontsize{8pt}{10pt}\selectfont
\psfrag{loc Aff}[B][B]{\shortstack{Locally\\non proportional\\linear dispersion}}
\psfrag{Affine}[l][l]{$k=\frac{\omega}{v_g}+k_0$}
\psfrag{phi}[l][l]{$\phi$}
\psfrag{k0}[b][b]{$k_0$}
\psfrag{w}[][]{$\omega$}
\psfrag{w0}[][]{$\omega_0$}
\psfrag{k=}[][]{}
\psfrag{Bl}[][]{\shortstack{$s(\nu)$\\band limited\\spectrum}}
\psfrag{D}[][]{$\Delta \omega$}
\psfrag{k}[][]{k}
\psfrag{dispersion}[][]{Dispersion plot}
\psfrag{linear}[l][l]{$k=\frac{\omega}{v_{\phi}}$}\includegraphics{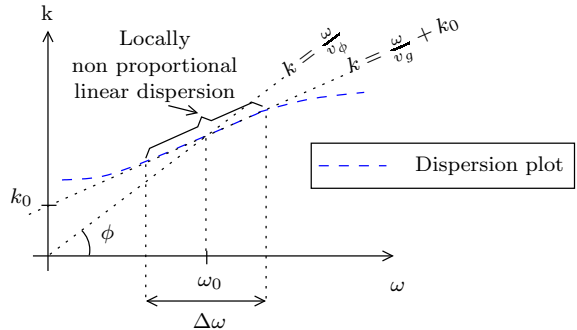}\protect\caption{Non proportional linear dispersion over a limited bandwidth.\label{fig:DispersionAndDelay}}
}
\end{figure}

\textcolor{black}{Nevertheless, if a limited bandwidth is acceptable
as in the case of modulated signals for telecommunication or radar
applications for instance, the delay line can be built using a media
where the dispersion is linear: $k=\frac{\omega}{v_{g}}+k_{0}$ and
non proportional ($k_{0}\neq0$) over the bandwidth as presented in
figure~\ref{fig:DispersionAndDelay} and where $v_{g}$ is the group
velocity defined as : }

\textcolor{black}{
\begin{eqnarray}
v_{g} & = & \frac{\text{d\ensuremath{\omega}}}{\text{d}k}\label{eq:GroupVelocity}
\end{eqnarray}
}

\textcolor{black}{In that case, and if additionally the amplitude
of the transmission coefficient remains constant in the bandwidth
as in the Shustter model, it has been shown by Brillouin~\cite{Brillouin1960}
that the propagation delay presented by the modulating signal is $t_{0}=\frac{L}{v_{g}}$
despite the phase velocity dispersion in the $\Delta\omega$ frequency
band.}

\textcolor{black}{Finally, to build up a delay line it is possible
to use a media presenting a non proportional linear dispersion from
0~Hz up to the maximal frequency of the propagating signal, consequently
where the phase and the group velocity are constant over this band.
It is also possible to use a media where the dispersion is non proportional
linear over a limited bandwidth and where only the group velocity
is constant over the considered band. The classical delay lines are
built according to the first method. In this paper, we focus on the
second method. }

\subsection{\textcolor{black}{Slowing down\label{sub:Slowing-down}}}

\begin{figure}[h]
\centering{}\textcolor{black}{\fontsize{8pt}{10pt}\selectfont
\psfrag{a}[l][l]{a)}
\psfrag{b}[l][l]{b)}
\psfrag{c}[l][l]{c)}
\psfrag{G}[r][r]{$G_{\text{dB}}\times \frac{v_0}{c}$}
\psfrag{Gu}[b][b]{dB/lu}Delete
\psfrag{Vg}[r][r]{$v_g / v_0 $}
\psfrag{Vgu}[B][b]{$\shortstack{no\\dimension}$}
\psfrag{k}[r][r]{{$\big(k(f)-k(f_0)\big) \frac{v_0}{c}$}}
\psfrag{ku}[b][b]{rad/lu}
\psfrag{g}[r][r]{$_{G_0}$}
\psfrag{0}[r][r]{0}
\psfrag{0.5}[r][r]{0.5}
\psfrag{1}[r][r]{1}
\psfrag{100}[r][r]{100}
\psfrag{-100}[r][r]{-100}
\psfrag{-200}[r][r]{-200}
\psfrag{-400}[r][r]{-400}
\psfrag{F}[][]{$0$}
\psfrag{+0.5}[][]{$+0.5$}
\psfrag{+1}[][]{$+1$}
\psfrag{+1.5}[][]{$+1.5$}
\psfrag{+2}[][]{$+2$}
\psfrag{-0.5}[][]{$-0.5$}
\psfrag{-1}[][]{$-1$}
\psfrag{-1.5}[][]{$-1.5$}
\psfrag{-2}[][]{$-2$}
\psfrag{D}[][]{$\Delta f$}
\psfrag{V}[][]{$\frac{\Delta v_{max}}{v_0}$}
\psfrag{GHz}[][]{GHz}
\psfrag{Axe}[][]{$(f-f_0)$}\includegraphics[width=10cm]{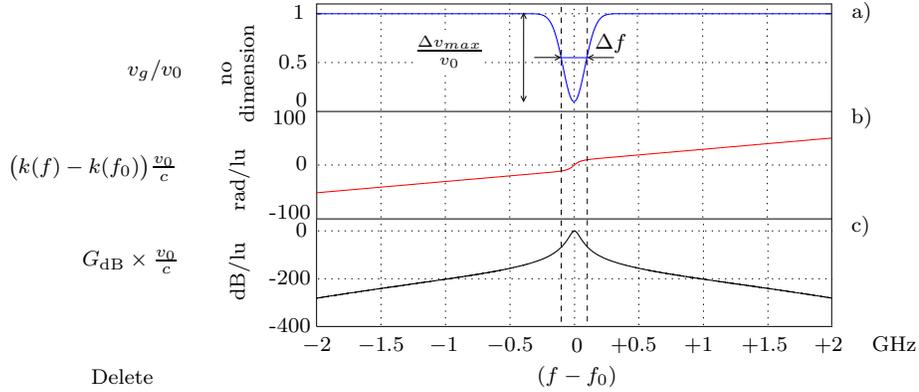}\protect\caption{Targeted velocity reduction and corresponding gain variation.\label{fig:Targeted-velocity-reduction}}
}
\end{figure}

\textcolor{black}{Considering the Kramers-Kronig relations~\cite{Kronig1926,Bohren2010},
group velocity variations can be obtained by modifying the amplitude
of the frequency response or transmission coefficient~\cite{Lucas2012}.
This is illustrated in figure~\ref{fig:Targeted-velocity-reduction}.
In this figure, curve (a) presents the relative targeted velocity
variation of the group velocity, where $v_{0}$ is the group velocity
outside the frequency band where the velocity variation occurs. The
shape of the velocity variation has been chosen arbitrarily. It is
nevertheless important that the velocity variation remains bounded
in frequency. In this figure, the frequencies are centered on $f_{0}$.
By using equation~\ref{eq:GroupVelocity} , one can calculate the
corresponding phase shift per length unit: 
\begin{eqnarray}
k(f) & = & \frac{2\pi}{v_{0}}\int\frac{1}{v_{n}(f)}{\rm d}f\label{eq:kvo/c}
\end{eqnarray}
}

\textcolor{black}{In this equation, $v_{n}(f)=\frac{v_{g}(f)}{v_{0}}$
which depends on $\Delta f$. figure~\ref{fig:Targeted-velocity-reduction}b
presents $\frac{v_{0}}{c}\Big(k(f)-k(f_{0})\Big)$ where c is the
speed of light in vacuum. This way, figure~\ref{fig:Targeted-velocity-reduction}b
is the same for all $f_{0}$. In this figure and later in the paper,
'lu' stands for length unit. It can be meters in continuous media
or a number of elementary elements in discrete media. }

\textcolor{black}{Considering reference~\cite{Lucas2012}~p~7~(equation\_18),
with ${\cal H}_{T}$ the Hilbert transform, one can obtain $G$(f)
the modulus of the frequency response of a one unit long lossless
line: }

\textcolor{black}{
\begin{eqnarray}
\frac{v_{0}}{c}\,\log(G(f)) & {\color{black}=} & {\color{black}-{\color{red}{\cal {\color{black}H_{T}}}{\color{black}\Bigg(\frac{_{v_{o}}}{c}\Big(k(f)-\frac{2\pi f}{v_{o}}\Big)\Bigg)}}}\label{eq:Hilbert(K)}
\end{eqnarray}
}

\textcolor{black}{$\frac{v_{0}}{c}\times G_{dB}$ is presented in
figure~\ref{fig:Targeted-velocity-reduction}c. Finally, one obtains
the required gain $G(f)$ that yields the expected velocity reduction
for an electromagnetic wave propagating in a given direction along
a line. Because the Hilbert transform of a constant is null, this
result can be shifted by any constant. In this figure we have chosen
to set the maximal value to zero dB. The data presented in this figure
have been obtained for a maximal relative velocity variation $\frac{\Delta v_{\text{max }}}{v_{0}}$
of 90~\% and a velocity reduction bandwidth at half depth $\Delta f$
of 100~MHz. }

\textcolor{black}{}

\paragraph*{\textcolor{black}{Losses }}

\textcolor{black}{In the precedent paragraph we have not considered
the losses. There are two main cases to consider. Whether the losses
are constant with the frequency as it is the case with resistive losses,
or they can be described with the constant loss angle approximation.
In both cases, they simply superimpose to the gain modulus variation
$G(f)$ calculated with equation~\ref{eq:Hilbert(K)}, without modifying
the phase shift and thus the group velocity. Details about this point
can be found in reference~\cite{Lucas2012}~pp~6-7.}

\subsection{\textcolor{black}{Parametric variations}}

\paragraph*{\textcolor{black}{Central frequency variation.}\protect \\
}

\textcolor{black}{Let us calculate the effect of a frequency shift
of the central frequency $f_{0}$. Let $f_{0}+f_{\text{offset}}$
be the new central frequency. $f_{\text{offset}}$ can be positive
or negative. The shifted velocity variation becomes : $v_{g}(f-f_{\text{offset}})$.
In turn, with equation~\ref{eq:kvo/c}, $k(f)$ becomes $k(f-f_{0}).$}

\textcolor{black}{As the Hilbert transform is the convolution with
$\frac{1}{j2\pi f}$ and as the convolution is associative and that
the Hilbert transform of a constant is null, the new required gain
variation $G_{\text{offset }}$ reads: 
\begin{eqnarray}
\frac{v_{0}}{c}\,\log(G_{\text{offset }}(f)) & = & -\frac{1}{j2\pi f}*\Bigg(\frac{2\pi}{v_{0}}k(f-f_{\text{offset }})-\frac{2\pi}{v_{0}}f\Bigg)\nonumber \\
 & = & -\frac{1}{j2\pi f}*\Bigg(\frac{2\pi}{v_{0}}k(f-f_{\text{offset }})-\frac{2\pi}{v_{0}}(f-f_{\text{offset }})+\underset{\text{constant}}{\underbrace{\frac{2\pi}{v_{0}}f_{\text{offset }}}}\Bigg)\nonumber \\
 & = & -\frac{1}{j2\pi f}*\Bigg(\Bigg[\frac{2\pi}{v_{0}}k(f)-\frac{2\pi}{v_{0}}(f)\Bigg]*\delta(f_{\text{offset }})\Bigg)\nonumber \\
 & = & \frac{v_{0}}{c}\,\log(G(f))*\delta(f_{\text{offset }})\label{eq:Goffset}
\end{eqnarray}
}

\textcolor{black}{In this equation $\delta$ denotes the Dirac distribution
and {*} is the convolution product. As one can see with this equation
shifting the targeted velocity variation simply shifts the required
gain. It does not change its amplitude.}

\textcolor{black}{As a conclusion to this paragraph, it is noticeable
that it is the bandwidth of the velocity reduction $\Delta f$ and
not the quality factor $\frac{f_{0}}{\Delta f}$ of the velocity reduction
that is relevant as a parameter which is quiet counter intuitive.
Therefore, in the parametric study that follows, it is the velocity
reduction bandwidth $\Delta f$ that will be considered as a parameter.}

\paragraph*{\textcolor{black}{Bandwidth $\Delta f$ and velocity reduction factor
$\frac{\Delta v_{\max}}{v_{o}}$ variations. }\protect \\
}

\textcolor{black}{The results of figure~\ref{fig:Targeted-velocity-reduction}
apply for any $v_{0}$ and central frequency $f_{0}$. However they
depend on the velocity reduction factor $\frac{\Delta v_{\max}}{v_{o}}$
and on the velocity reduction bandwidth $\Delta f$.}

\begin{figure}[h]
\centering{}\textcolor{black}{\fontsize{8pt}{10pt}\selectfont
\psfrag{G}[r][r]{$G_{\text{dB}}\times \frac{v_0}{c}$}
\psfrag{Gu}[b][b]{dB/lu}
\psfrag{Vg}[][]{$v_g / v_0 $}
\psfrag{Vgu}[b][b]{$\shortstack{no\\dimension}$}
\psfrag{k}[r][r]{{$\big(k(f)-k(f_0)\big) \frac{v_0}{c}$}}
\psfrag{ku}[][]{$\text{rad}/\text{lu}$}
\psfrag{0}[r][r]{0}
\psfrag{100}[r][r]{100}
\psfrag{-100}[r][r]{-100}
\psfrag{0.5}[r][r]{+0.5}
\psfrag{1}[r][r]{+1}
\psfrag{-200}[r][r]{-200}
\psfrag{-400}[r][r]{-400}
\psfrag{1.5}[][]{+1.5}
\psfrag{2}[][]{+2}
\psfrag{-0.5}[][]{-0.5}
\psfrag{-1}[][]{-1}
\psfrag{-1.5}[][]{-1.5}
\psfrag{-2}[][]{-2}
\psfrag{GHz}[][]{GHz}
\psfrag{Axe}[][]{$(f-f_0)$}\includegraphics[width=10cm]{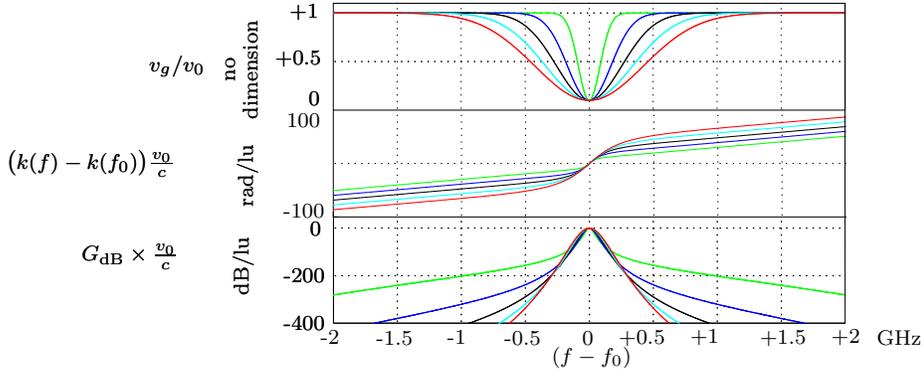}\protect\caption{Influence of the bandwidth of the velocity reduction.\label{fig:InfluenceBandwidth}}
}
\end{figure}

\textcolor{black}{We have computed the required gain $G(f)$ for various
values of $\Delta f$ as presented in figure~\ref{fig:InfluenceBandwidth}.}

\textcolor{black}{}

\begin{figure}[h]
\centering{}\textcolor{black}{\fontsize{8pt}{10pt}\selectfont
\psfrag{0}[][]{0}
\psfrag{5}[][]{5}
\psfrag{10}[][]{10}
\psfrag{20}[][]{20}
\psfrag{30}[][]{30}
\psfrag{40}[][]{40}
\psfrag{50}[][]{50}
\psfrag{60}[][]{60}
\psfrag{70}[][]{70}
\psfrag{80}[][]{80}
\psfrag{90}[][]{90}
\psfrag{100}[][]{100}
\psfrag{120}[][]{120}
\psfrag{140}[][]{140}
\psfrag{0}[][]{0}
\psfrag{50}[][]{50}
\psfrag{100}[][]{100}
\psfrag{150}[][]{150}
\psfrag{200}[][]{200}
\psfrag{250}[][]{250}
\psfrag{300}[][]{300}
\psfrag{350}[][]{350}
\psfrag{400}[][]{400}
\psfrag{450}[][]{450}
\psfrag{500}[][]{500}
\psfrag{Width}[][]{$\Delta f$ (MHz)}
\psfrag{S21db}[c][c]{$G_{\text{dB}} \times \frac{v_0}{c}$ (dB/lu)}
\psfrag{DeltaV/V}[][]{$ \frac{\Delta v_{max}}{v_0}$\%}
\psfrag{a}[][]{a)}
\psfrag{b}[][]{b)}
\psfrag{XFmax}[B][b]{$\times f_{max}$}\includegraphics[scale=0.5]{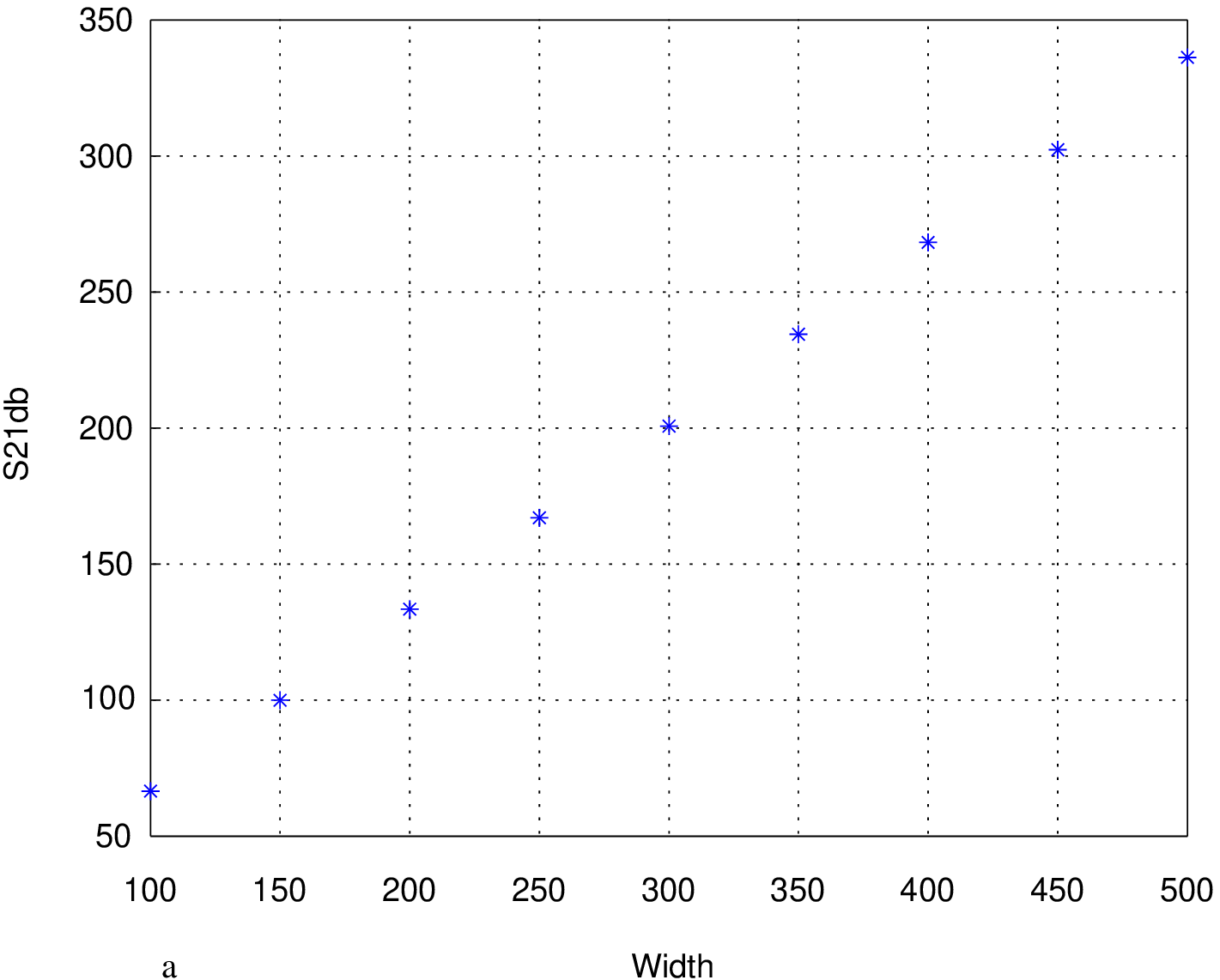}\hspace{8 pt}\includegraphics[scale=0.5]{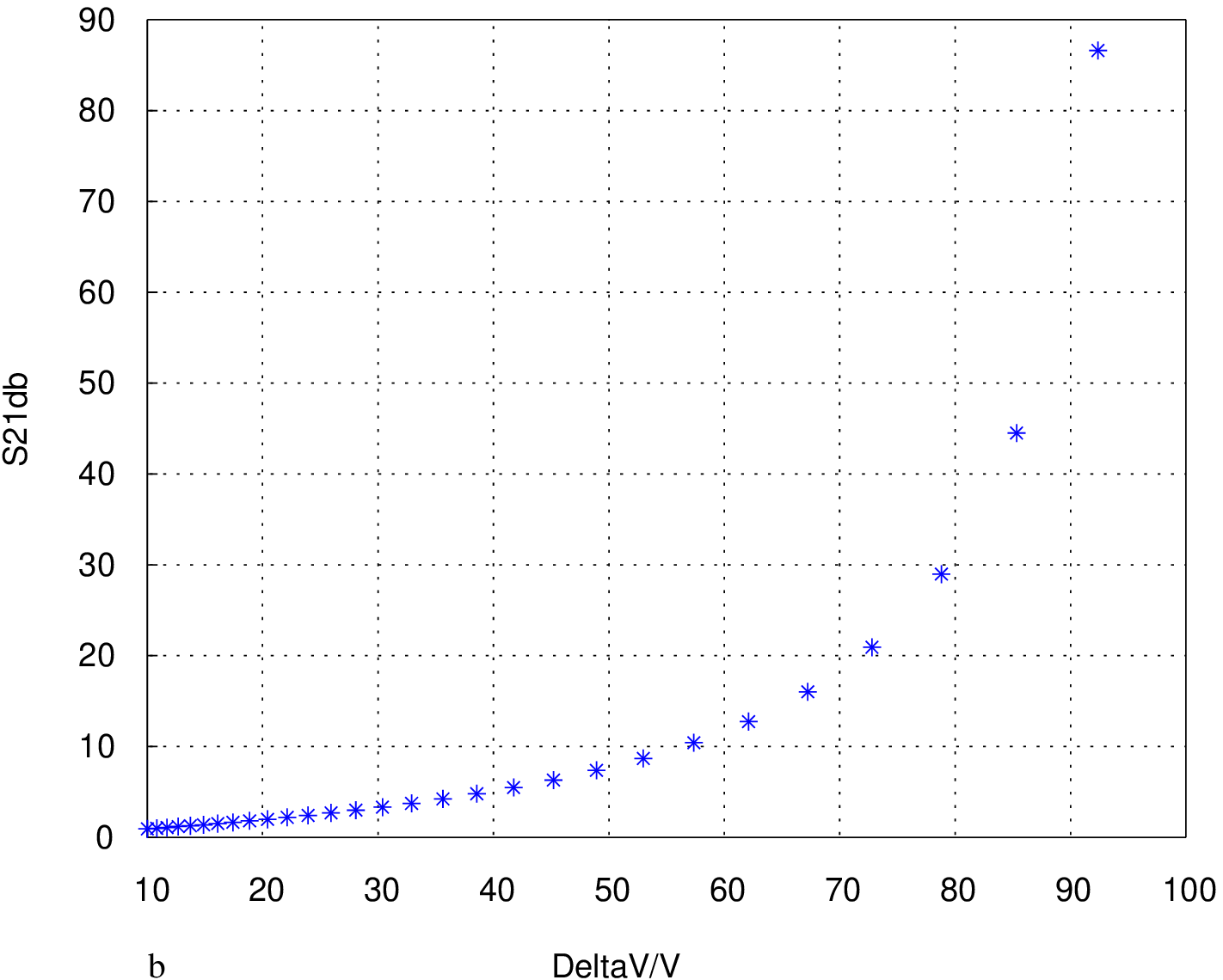}
\protect\caption{Influences of the bandwidth and relative velocity depth variation.\label{fig:ParametricGain}}
}
\end{figure}

\textcolor{black}{By doing so, and proceeding in the same way for
various values of $\frac{\Delta v_{\max}}{v_{o}}$ , one obtains the
plots presented in figure~\ref{fig:ParametricGain}. In figure~\ref{fig:ParametricGain}a
the variation of the required maximal value for $G(f)$ as a function
of the bandwidth is presented, when figure~\ref{fig:ParametricGain}b
presents the same parameter versus the relative velocity variation. }

\textcolor{black}{}

\textcolor{black}{Considering figure~\ref{fig:ParametricGain}b,
one can see that it becomes very costly in terms of gain to reduce
the group velocity beyond 90 \%. In this work we targeted a velocity
reduction of 90~\%. figure~\ref{fig:ParametricGain}a shows that
the required gain is linear in dB. It actually increases exponentially.
In section \ref{sec:Delay-lines_General}, we have seen that it is
important to have a non proportional linear dispersion in order to
allow band limited signal to propagate at the group velocity. One
way to obtain this non proportional linear dispersion consists in
working with a sufficiently large bandwidth compared to that of the
transmitted signal.}

\subsection{\textcolor{black}{Applications. \label{sub:Applications}}}

\paragraph*{\textcolor{black}{Line approach:}}

\textcolor{black}{By targeting a group velocity profile, one calculates
naturally the characteristics a line should present: the wave vector
and the gain per length unit. To obtain a given delay $\Delta T$
one simply sets the length $L$ of the line to $L=v_{g}.\Delta T$.
This is the great point about the line approach. Nevertheless, the
precedent results can be applied only if the line segment is properly
loaded so that there is only the incident wave to consider. It should
be noted that setting the group velocity sets as well the characteristic
impedance which as to be taken into account when building a prototype.}

\paragraph*{\textcolor{black}{Filter approach:}}

\textcolor{black}{A system which exhibits the gain of figure~\ref{fig:Targeted-velocity-reduction}c
is a filter. It is indeed possible to obtain constant delay over a
bandwidth with high order Bessel's filters. The order of the filter
is related to the delay expected. It is always possible anyhow to
split high order filter into a daisy chain of first and second order
filters. It consists in fact in building a metamaterial line segment.
In this paper we chose to use the line approach because it closely
fits the metamaterial theoretical approach. It is particularly true
in so far the impedance matching issue is concerned.}

\paragraph*{\textcolor{black}{Example of application:}}

\textcolor{black}{In this paper, we consider as an example a telecommunication
application around 2~GHz such as the DCS 1800 (Digital Cellular System
) which is compliant with the GSM (Global System for Mobile Communications)
protocol. The transmitted signals are narrow band to allow numerous
transmission channels. A 100~MHz bandwidth allows to obtain a sufficiently
non proportional linear dispersion around the central frequency for
the typical channel width of the GSM protocol (200 kHz). Lets us use
the results of the precedent section to estimate the required gain
per length unit to obtain a group velocity reduction of 90\%. With
$v_{g}\approx\frac{2}{3}c$ which is the group velocity in a 50~$\Omega$
coaxial cable, according to figure~\ref{fig:ParametricGain}a, one
has: $\frac{v_{g}}{v_{0}}G_{o}\approx-\text{66\,\text{dB/lu}\ensuremath{\rightarrow}}G_{0}\approx-100\,\text{dB/lu}$.
This corresponds to a very steep gain at the borders of the band.
This can typically be achieved by the exponential attenuation between
two forbidden band gaps in a metamaterial \cite{Trytiakov2007} or
by using a resonant design as in reference \cite{Lucyszyn1995}. This
work is focused on the first technique because metamaterial allows
to easily control $\Delta f$ which is the relevant parameter of slow
line design.}

\section{\textcolor{black}{1D slow metamaterial lines}}

\begin{figure}[h]
\centering{}\textcolor{black}{\fontsize{8pt}{10pt}\selectfont
\psfrag{l}[][]{$\ell$}
\psfrag{n}[][]{n}
\psfrag{n+1}[][]{n+1}
\psfrag{n-1}[][]{n-1}
\psfrag{Comp}[][]{Components}
\psfrag{segments}[][]{\shortstack{Line\\segment}}
\psfrag{C1}[][]{$C_1$}
\psfrag{L1}[][]{$L_1$}
\psfrag{C2}[][]{$C_2$}
\psfrag{a}[l][l]{a)}
\psfrag{b}[l][l]{b)}
\psfrag{E}[][]{Elementary cell}\includegraphics{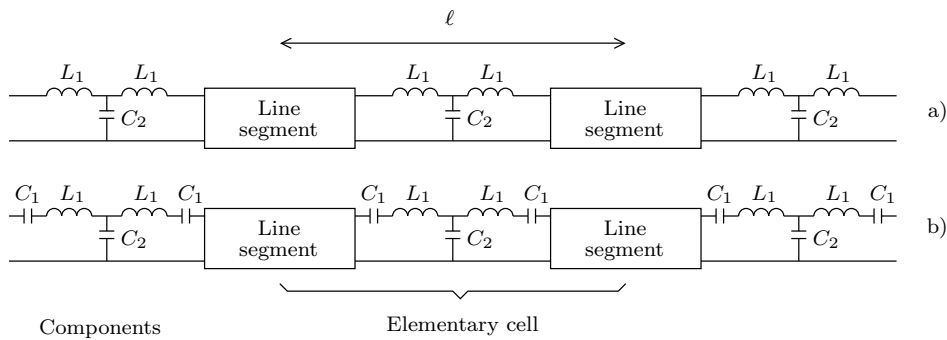}\protect\caption{1D Metamaterial slow lines: a) Classical LC line b) Band Gap Slow
line\label{fig:1DMetamaterial}}
}
\end{figure}

\textcolor{black}{Electrical delay lines are classically build by
cascading L-C cells as presented in figure~\ref{fig:1DMetamaterial}a.
Such lines are in fact a 1D metamaterial line. 1D metamaterial are
made of elementary cells made of lumped components which, from an
electrical point of view, are two ports circuits. Practically, the
elementary cells and components are connected using line segments.
Such material behave differently whether the propagation in the line
segment can be neglected in front of the phase shift imposed by the
lumped components or not. This is often referred as the sub-wavelength
condition in the literature. In terms of material science, the 1D
metamaterial line will be used in its first Brillouin zone~\cite{Brillouin1956}
where the elementary cell length $\ell$ is such that the line appears
homogeneous to the electromagnetic wave it carries. This condition
can be stated for instance as: 
\begin{eqnarray}
\ell & < & \frac{\lambda}{10}\label{eq:MetaCondition}
\end{eqnarray}
where $\lambda$ is the wave length of the carried wave. }

\textcolor{black}{}

\textcolor{black}{Under the sub-wavelength condition, only the lumped
components are taken into account. Using Kirchhoff laws and Floquet's
theorem, the phase shift per cell for an infinitely long line can
be calculated. Equation \ref{eq:CosTethaLC} \cite{Eleftheriades2005,Caloz2006}
presents the theoretical dispersion for the classical LC structure
of figure~\ref{fig:1DMetamaterial}a. 
\begin{eqnarray}
\cos(\theta) & = & 1-L_{1}C_{2}\omega^{2}\label{eq:CosTethaLC}
\end{eqnarray}
}

\begin{figure}[h]
\centering{}\textcolor{black}{\fontsize{8pt}{10pt}\selectfont
\psfrag{A}[][]{$G_{\text{dB}}$ dB/Cells}
\psfrag{f}[][]{GHz}
\psfrag{F}[][]{Frequency}
\psfrag{0}[][]{0}
\psfrag{10}[][]{10}
\psfrag{-10}[][]{-10}
\psfrag{-20}[][]{-20}
\psfrag{-30}[][]{-30}
\psfrag{-40}[][]{-40}
\psfrag{-50}[][]{-50}
\psfrag{0.5}[][]{0.5}
\psfrag{1}[][]{1}
\psfrag{1.5}[][]{1.5}
\psfrag{2}[][]{2}
\psfrag{2.5}[][]{2.5}
\psfrag{3}[][]{3}
\psfrag{3.5}[][]{3.5}
\psfrag{4}[][]{4}
\psfrag{w1}[r][r]{$f_1$}
\psfrag{w2}[l][l]{$f_2$}
\psfrag{wc}[][]{$\omega_2=\frac{1}{\sqrt{L_1 C_2}}$}
\psfrag{Vg}[][]{$V_g$ Cells/s}
\psfrag{c000}[][]{$10^9$}
\psfrag{c010}[][]{$10^8$}
\psfrag{c100}[][]{$10^7$}\includegraphics[width=8cm]{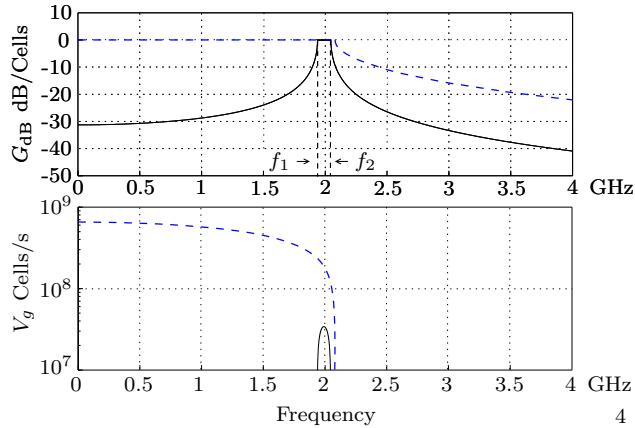}\ref{eq:Goffset}\protect\caption{1D slow line metamaterial behavior\label{fig:Behavior}}
}
\end{figure}

\textcolor{black}{In this equation, $\theta$ is the phase shift per
cell which can be complex. In sub wavelength condition, the length
of the line $L$, as defined in section~\ref{sec:Delay-lines_General},
is the number of cells, and one has: $v_{g}={\rm d}\theta/{\rm d\omega}$
in cells per seconds. With $\ell$ the length of a cell it comes:
$v_{g}={\rm d}\theta/{\rm d\omega}.\ell$ in m/s. Considering the
dispersion relation~\ref{eq:CosTethaLC}, waves are evanescent whenever
$|\cos(\theta)|>1$. Concurrently G(f) goes very low. With equation~\ref{eq:CosTethaLC},
one can calculate the corresponding modulus $G(f)=\Big|e^{-j\theta}\Big|$
of the frequency response for one cell of an infinitely long line.
G(f) for the classical LC structure is presented in dotted blue line
in figure~\ref{fig:Behavior} along with the corresponding velocity.
This kind of line is a low-pass structure. To make it work at 2 Ghz
as proposed in section \ref{sub:Applications} implies a very large
bandwidth which is not favorable in so far velocity reduction is concerned
as it is shown in figure~\ref{fig:ParametricGain}a. In order to
reduce this bandwidth we inserted a serial capacitor in the line to
cut the low frequencies. The new line structure is presented in figure~\ref{fig:1DMetamaterial}b.
Equation~\ref{eq:CosTethaLC} becomes \cite{Eleftheriades2005,Caloz2006}:}

\textcolor{black}{
\begin{eqnarray}
\cos(\theta) & = & 1+\frac{C_{2}}{C_{1}}-L_{1}C_{2}\omega^{2}\label{eq:CosTethaBG}\\
f_{1} & = & \frac{1}{2\pi\sqrt{L_{1}C_{1}}}\label{eq:f1}\\
f_{2} & = & \frac{1}{2\pi}\sqrt{\frac{1}{L_{1}C_{1}}+\frac{2}{L_{1}C_{2}}}\label{eq:f2}
\end{eqnarray}
}

\textcolor{black}{One can see that the introduction the serial capacitor
shifts up $\cos(\theta)$, thus creating a new bandgap that starts
at zero Hertz. In DC condition, $\cos(\theta)>1$ because of the capacitor
$C_{1}$. When the angular frequency gets greater than $f_{1}$, normal
propagation occurs till $\cos(\theta)<1$ which corresponds to $f_{2}$
. This behavior is summarized by the black solid curves in figure~\ref{fig:Behavior}.}

\textcolor{black}{}

\textcolor{black}{This figure shows that there is, in the metamaterial,
a propagation band flanked by two forbidden bands on each side. This
reduces the band width and makes the slopes more steep. As expected
from section~\ref{sub:Slowing-down}, it reduces as well the group
velocity in the pass-band. }

\textcolor{black}{It is possible to place $f_{1}$ and $f_{2}$ independently
on the frequency axis. This allows to obtain the low propagation velocity
presented in figure~\ref{fig:Behavior} at any frequency in sub-wavelength
condition at the cost of the characteristic impedance. Indeed considering
an infinity long line, it is also possible to calculate the characteristic
impedance : 
\begin{eqnarray}
Z_{c}^{2} & = & 2Z_{1}Z_{2}+Z_{1}^{2}\label{eq:Zc}
\end{eqnarray}
}

\textcolor{black}{where $Z_{1}=\frac{1}{j\omega C_{1}}+jL_{1}\omega$
and $Z_{2}=\frac{1}{jC\omega_{2}}$. Setting $v_{g}$ this is to says
$f_{1}$ and $f_{2}$ consequently sets $Z_{c}$.}

\textcolor{black}{Finally, one has to find a compromise between velocity
reduction and line impedance matching.}

\textcolor{black}{The velocity profile obtained in figure~\ref{fig:Behavior}
is not the one considered in section~\ref{sub:Slowing-down}. Indeed
in this introductory section a media were the waves propagate for
all frequencies was considered. In a band gap metamaterial, the waves
only propagate in the pass-bands. Outside of this pass-bands the group
velocity is infinite and has no particular physical signification
and the wave vector modulus $k$ is constant with $k=n.\pi$ where
n is an integer. It only varies in the pass-band. In both cases, the
variation of the wave vector modulus $k$ in theses bands is positive
as in the case considered in section~\ref{sub:Slowing-down}. This
of course leads in both cases to a corresponding required gain variation
shaped as a pass-band. Further more in the case of a band gap metamaterial
one can approximate the group velocity as: $v_{g}\approx\frac{\Delta\omega}{\Delta k}=\frac{2\pi\Delta F}{\pi}=2\Delta F$
in cells per seconds. This further illustrate the general results
of equation~\ref{eq:Goffset} which states that $\Delta F$ is the
relevant parameter. Finally as band gap metamaterial makes it possible
to control $\Delta F$, they are an excellent way for building slow
lines. }

\section{\textcolor{black}{Tuning the slow line }}

\textcolor{black}{In this work, the targeted frequency is about 2~GHz
(CF section~\ref{sub:Applications}), therefore equation~\ref{eq:MetaCondition}
yields $\ell<\text{1 cm }$ when using microstrip lines over a PTFE
woven glass substrate. This condition can be respected using classical
hybrid technology~\cite{Geron2012}. Consequently, the 1D metamaterial
components values are calculated in a first time without considering
the line segment using equations~\ref{eq:f1} and \ref{eq:f2}. The
design is then post corrected to take into account the inherent serial
inductance of the line segments, which adds to $L_{1}$, and parallel
capacitance which adds to $C_{2}$. They are both assumed to be small
to bring only minor corrections.}

\textcolor{black}{In the realization presented hereafter we have chosen
to favor velocity reduction over characteristic impedance. One can
find in Table~\ref{tab:Values} the values calculated to obtain a
bandwidth of about 100 MHz around 2~Ghz. In the same table, one will
find as well, the values corrected by simulating~\cite{ADS2012}
a line of 10 cells of the metamaterial. It practically corresponds
to a 7~cm long line taking into account the line segments used to
practically build the line.}

\textcolor{black}{}
\begin{table}[h]
\centering{}\textcolor{black}{}%
\begin{tabular}{|c|>{\raggedright}p{4cm}|>{\centering}p{3cm}|}
\hline 
 & \textcolor{black}{Theoretical values without line segment } & \textcolor{black}{Corrected values with line segment}\tabularnewline
\hline 
\hline 
\textcolor{black}{$L_{1}$} & \textcolor{black}{4.5~nH} & \textcolor{black}{3.4~nH}\tabularnewline
\hline 
\textcolor{black}{$C_{1}$} & \textcolor{black}{1.5~pF} & \textcolor{black}{1.4~pF}\tabularnewline
\hline 
\textcolor{black}{$C_{2}$} & \textcolor{black}{26~pF} & \textcolor{black}{21~pF}\tabularnewline
\hline 
\end{tabular}\textcolor{black}{\protect\caption{Values calculated using equation \ref{eq:CosTethaBG} and post corrected
by simulation for the 1D slow metamaterial line.\label{tab:Values}}
}
\end{table}

\textcolor{black}{The obtained line presents a low characteristic
impedance of about 3~$\Omega$ which makes it necessary to use quarter
wave line segments at its input and output in order to avoid reflections.
As these quarter wave line segments are not slow line segments, the
multiple reflections they induce are short lived compared to the propagation
time in the line itself and do not perturb the line propagation of
the signal. This point will be verified with the transient measurements
of section~\ref{sub:Performances}.}

\textcolor{black}{}
\begin{figure}[h]
\begin{centering}
\textcolor{black}{\fontsize{8pt}{10pt}\selectfont
\psfrag{db}[][]{$\vert \text S_{21}\vert $ dB}
\psfrag{Ghz}[][]{GHz}
\psfrag{_000}[r][r]{0}
\psfrag{_50}[r][r]{-50}
\psfrag{_150}[r][r]{-150}
\psfrag{_250}[r][r]{-250}
\psfrag{0}[][]{0}
\psfrag{2}[][]{2}
\psfrag{4}[][]{4}
\psfrag{6}[][]{6}
\psfrag{8}[][]{8}
\psfrag{10}[][]{10}
\psfrag{12}[][]{12}
\psfrag{14}[][]{14}\psfrag{G}[][]{GHz}
\psfrag{ns}[][]{ns}
\psfrag{1.8}[][]{1.8}
\psfrag{1.9}[][]{1.9}
\psfrag{2.0}[][]{2.0}
\psfrag{2.1}[][]{2.1}
\psfrag{2.2}[][]{2.2}
\psfrag{3.}[][]{2.5}
\psfrag{2.}[][]{2}
\psfrag{1.}[][]{1}
\psfrag{1.5}[][]{1.5}
\psfrag{Arbitrary}[][]{Arbitrary Units}
\psfrag{a}[l][l]{a) Forbidden bands overview}
\psfrag{c}[l][l]{c)  Group delay}
\psfrag{b}[l][l]{b)  arg(S$_{21}$)}
\psfrag{000}[r][r]{0}
\psfrag{200}[r][r]{200}
\psfrag{150}[r][r]{150}
\psfrag{100}[r][r]{100}
\psfrag{050}[r][r]{50}
\psfrag{025}[c][c]{$\approx$ 29 ns}\includegraphics[width=15cm]{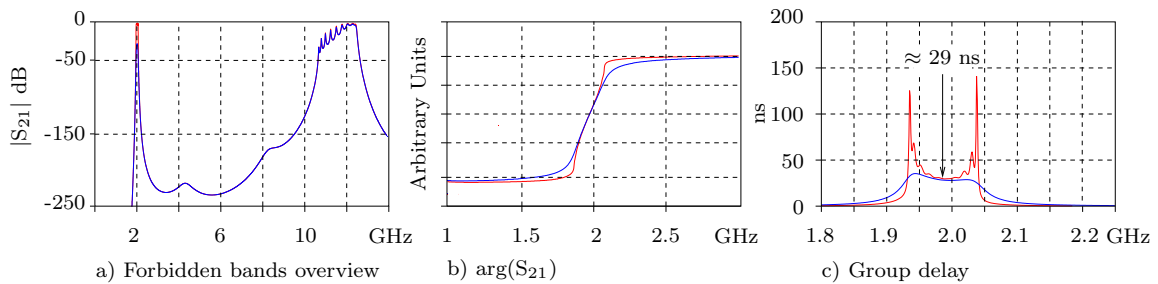}}
\par\end{centering}

\centering{}\textcolor{black}{\protect\caption{Simulated slow line forbidden bands and performances. Without losses
(red). With losses (blue)\label{fig:SlowLineSimPerformances}}
}
\end{figure}

\textcolor{black}{figure \ref{fig:SlowLineSimPerformances}a presents
an overview of the behavior of the transmission for a 10 cells line.
It exhibits a narrow passband with very steep slopes around 2~GHz
as required to significantly reduce the group velocity. When the frequency
goes higher a second passband appears. It corresponds to the second
Brillouin zone when the ``classical'' propagation cannot be neglected
because the condition of equation~\ref{eq:MetaCondition} is no longer
respected. The phenomenon that creates this pass-band is the classical
Bragg effect. figure~\ref{fig:SlowLineSimPerformances}b presents
the shape of the argument of the transmission coefficient, which shows
that there is indeed a strong variation of the group velocity in the
considered passband. Finally, figure~\ref{fig:SlowLineSimPerformances}c
presents the corresponding group delay. The 29~ns at the center of
the band are to be compared to the $\approx$~0.35~ns that would
be obtained in a classical 12~cm long microstrip line, which is the
length of the prototype.}

\section{\textcolor{black}{Building and characterizing the slow line}}

\subsection{\textcolor{black}{Realization }}

\begin{figure}[h]
\begin{centering}
\textcolor{black}{\fontsize{8pt}{10pt}\selectfont
\psfrag{12cm}[t][t]{\textcolor{white}{$\approx$ 7 cm}}
\psfrag{cell}[l][l]{\shortstack[r]{Cell :  Lumped components connected\\using microstrip line segments}}
\psfrag{Adaptation}[l][l]{Quarter wave adaptation line}
\psfrag{Ground}[l][l]{\shortstack{Top side ground plane and\\vias to bottom side ground plane}}
\psfrag{Substrat}[lb][lb]{\shortstack{PTFE Woven Glass \\Substrate}}
\psfrag{C2}[][]{$C_2$}
\psfrag{C1}[][]{$C_1$}
\psfrag{2L1}[][]{$2\times L_1$}\includegraphics[width=13cm]{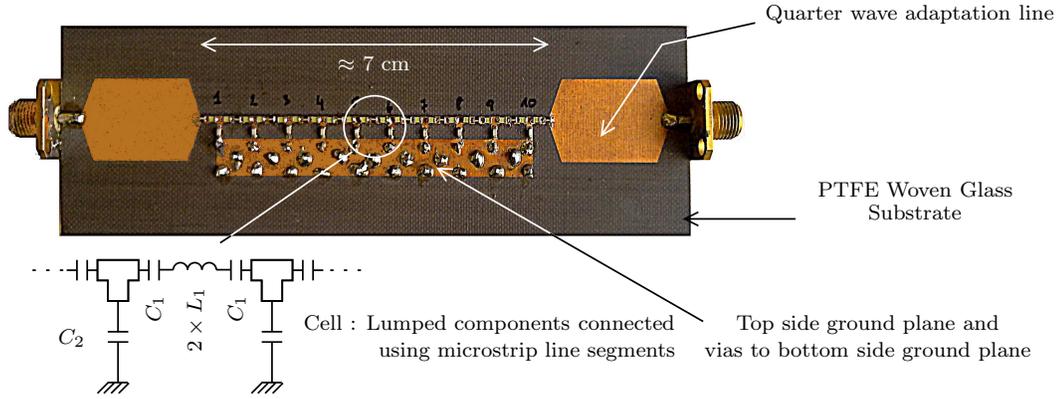}}
\par\end{centering}

\textcolor{black}{\protect\caption{The slow line prototype.\label{fig:Slow-line-prototype}}
}
\end{figure}

\textcolor{black}{The line calculated and simulated has been realized
and is presented in figure~\ref{fig:Slow-line-prototype}. In this
figure, one can see the 1D metamaterial line between its two quarter
wave length tapered lines. These lines realize the impedance matching
between the 50~$\Omega$ connectors and the 3~$\Omega$ metamaterial
line. The 12~$\Omega$ required for the quarter wave length lines
makes them very large compared to the metamaterial line, which is
the reason why they gradually increase and decrease at each end to
avoid a potentially mismatching step. One can see as well that the
connection to the ground is realized through a top side ground plane
whose equipotentiality is ensured by multiple vias to the bottom ground
plane. The zoom on one of the cells details the practical implementation
the line. One may notice the use of a $2L_{1}$ inductor and two $C_{1}$
capacitors which allows to maintain the cell symmetry while reducing
the number of lumped components.}

\textcolor{black}{}

\subsection{\textcolor{black}{Performances\label{sub:Performances}}}

\textcolor{black}{We have firstly measured the value of the group
velocity and group delay for the prototype. This measurement was simply
realized using a Vector Network Analyzer (VNA) to measure the transmission
coefficient $\text{S}_{21}$. This corresponds to a steady state measurement.
The corresponding group delay $GD$ and group velocity $v_{g}$ can
be obtained with equations \ref{eq:GD} and \ref{eq:VG}:
\begin{eqnarray}
GD & = & 2\pi\Bigg(\frac{\text{dS}_{21}}{\text{d}f}\Bigg)^{-1}\label{eq:GD}\\
v_{g} & = & L/GD\label{eq:VG}
\end{eqnarray}
}

\textcolor{black}{where L=12~cm is the total length of the prototype. }

\begin{figure}[h]
\begin{centering}
\textcolor{black}{\fontsize{8pt}{10pt}\selectfont
\psfrag{dB}[][]{$\vert \text S_{21}\vert $ dB}
\psfrag{GHz}[][]{GHz}
\psfrag{ns}[][]{ns}
\psfrag{linear}[][]{\shortstack{linear\\regression}}
\psfrag{38}[][]{-38 dB}
\psfrag{0}[][]{0}
\psfrag{1}[][]{1}
\psfrag{2}[][]{2}
\psfrag{3}[][]{3}
\psfrag{4}[][]{4}
\psfrag{5}[][]{5}
\psfrag{6}[][]{6}
\psfrag{7}[][]{7}
\psfrag{-40}[][]{-40}
\psfrag{-20}[][]{-20}
\psfrag{0}[][]{0}
\psfrag{20}[][]{20}
\psfrag{40}[][]{40}
\psfrag{60}[][]{60}
\psfrag{1.9}[][]{1.90}
\psfrag{1.95}[][]{1.95}
\psfrag{2.00}[][]{2.00}
\psfrag{2.05}[][]{2.05}
\psfrag{-75}[][]{-75}
\psfrag{-70}[][]{-70}
\psfrag{-65}[][]{-65}
\psfrag{-60}[][]{-60}
\psfrag{-55}[][]{-55}
\psfrag{-50}[][]{-50}
\psfrag{-45}[][]{-45}
\psfrag{-40}[][]{-40}
\psfrag{-35}[][]{-35}
\psfrag{rad}[][]{$\times \pi$ rad}
\psfrag{a}[l][l]{a) $\vert $S$_{21}$ $\vert $}
\psfrag{c}[l][l]{c)  Group delay}
\psfrag{b}[l][l]{b)  arg(S$_{21}$)}\includegraphics[width=15cm]{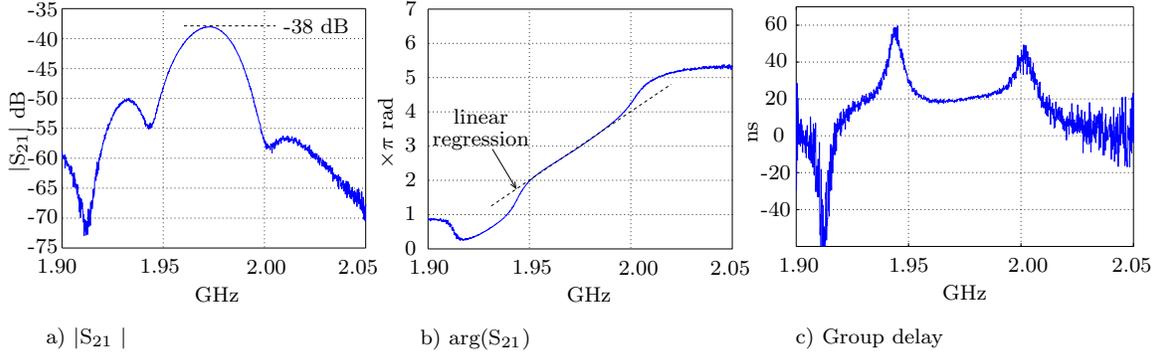}}
\par\end{centering}

\centering{}\textcolor{black}{\protect\caption{Steady state Measurement of the slow line performances.\label{fig:SlowLineMeasPerformances}}
}
\end{figure}

\textcolor{black}{The results are presented in figure~\ref{fig:SlowLineMeasPerformances}.
They are narrow band compared to the simulation which allows to properly
calibrate the VNA. One can see that as long as the amplitude of $|\text{S}_{21}|$
is sufficiently large, they are comparable to the simulation results
presented in figure~\ref{fig:SlowLineSimPerformances}. By measuring
the group velocity using a linear regression as presented in figure
\ref{fig:SlowLineMeasPerformances}a, the velocity reduction ratio
is 35 as presented in Table~\ref{tab:Peformances}.}

\textcolor{black}{}

\textcolor{black}{The attenuation (-38 dB) is quiet strong and comparable
to the simulation where the measured the losses of each kind of components
was used. We can conclude that it is is mostly due to the components.
At this level, these losses prevent of course to use such a line for
industrial applications. Nevertheless it allows to validate the concept.
Though being high, the losses do not prevent to measure the effective
data transmission ability of the slow line. }

\textcolor{black}{}
\begin{table}[h]
\begin{centering}
\textcolor{black}{}%
\begin{tabular}{|>{\centering}m{4cm}|c|>{\centering}m{2cm}|>{\centering}m{1.5cm}|>{\centering}m{2cm}|>{\centering}m{1.5cm}|}
\hline 
\multirow{1}{4cm}[1cm]{} & \textcolor{black}{$V_{g_{\text{ref }}}$} & \textcolor{black}{$V_{g_{\text{Slow }}}$} & \textcolor{black}{$\frac{V_{g_{\text{ref }}}}{V_{g_{\text{Slow }}}}$} & \textcolor{black}{Reflection}

\textcolor{black}{($\vert S_{11}\vert$)} & \textcolor{black}{Losses}\tabularnewline
\hline 
\hline 
\textcolor{black}{Simulated} & \textcolor{black}{2.06 $10^{8}$ m/s} & \textcolor{black}{4.20 $10^{6}$ m/s} & \textcolor{black}{48} & \textcolor{black}{-15 dB} & \textcolor{black}{-29 dB}\tabularnewline
\hline 
\textcolor{black}{Steady State measurement}

\textcolor{black}{(via VNA $S_{21}$)} & \textcolor{black}{2.1 $10^{8}$ m/s} & \textcolor{black}{5.99 $10^{6}$ m/s} & \textcolor{black}{35} & \textcolor{black}{-10 dB} & \textcolor{black}{-38 dB}\tabularnewline
\hline 
\textcolor{black}{Transient measurement}

\textcolor{black}{(Zero span SPA) } & \textcolor{black}{2.14 $10^{8}$ m/s} & \textcolor{black}{5.84 $10^{6}$ m/s} & \textcolor{black}{37} & \textcolor{black}{{*}{*}} & \textcolor{black}{-40 dB}\tabularnewline
\hline 
\end{tabular}
\par\end{centering}

\centering{}\textcolor{black}{{*}{*} = Not measured \protect\caption{Summary of the performances of the slow line.\label{tab:Peformances}}
}
\end{table}

\begin{figure}[h]
\centering{}\textcolor{black}{\fontsize{8pt}{10pt}\selectfont
\psfrag{MOD}[][]{\shortstack{2 GFSK\\modulator}}
\psfrag{Ref}[][]{\shortstack{Reference\\line}}
\psfrag{Slow}[][]{\shortstack{Slow\\line}}
\psfrag{Synchro}[][]{Synchronization}
\psfrag{Zero Span SPA}[][]{Zero Span spectrum analyser}
\psfrag{dt}[][]{dt}
\psfrag{t}[][]{t}
\psfrag{dB}[][]{dB}
\psfrag{-40}[][]{-40}
\psfrag{0}[][]{0}
\psfrag{Symrate}[l][l]{Symbol rate: 5 MHz}
\psfrag{DF}[l][l]{Modulation: $\frac{\Delta f}{f_m}=0.5$}
\psfrag{BT}[lt][lt]{Filtering: BT=0.3}
\psfrag{Power}[l][l]{Power ramping}
\psfrag{F}[l][l]{$f_0$}
\psfrag{Modulated}[][]{\shortstack{Narrow band\\signal}}
\psfrag{Modulation}[b][b]{\shortstack{Modulation\\Characteristics}}
\psfrag{lo}[r][r]{Local Oscillator}
\psfrag{trame}[r][r]{Data frame}\includegraphics{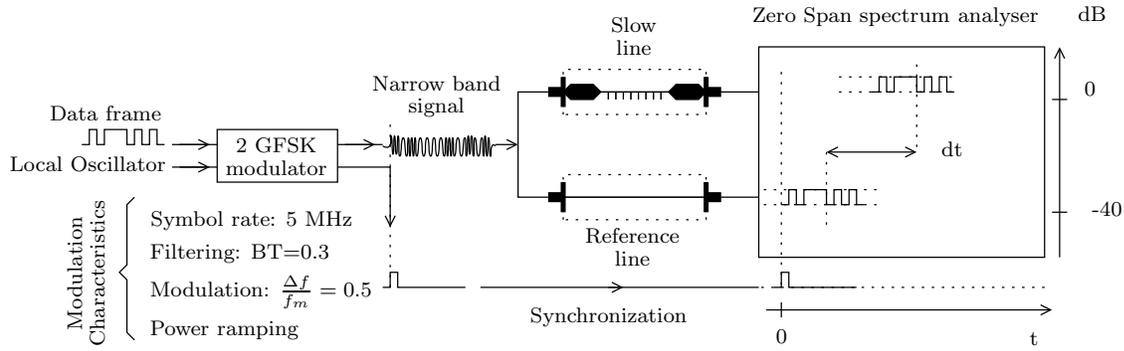}\protect\caption{Experimental signal velocity measurement setup.\label{fig:Setup}}
}
\end{figure}

\textcolor{black}{We have verified the ability of the built up slow
line to effectively transmit data at slow velocity. To do so, we measured
the effective signal velocity by comparing the transit time of a Gaussian
filtered CPFSK ( Gaussian Frequency Shift Keying )~\cite{Jordan1989}
modulated data frame in the slow line to the one in a reference microstrip
line of the same length. By doing so, we are sure to effectively measure
the information velocity, which would have not been fully the case
if we had used a Gaussian modulated signal for instance~\cite{Sauter2002}.
The experimental setup is presented in figure~\ref{fig:Setup}.}

\textcolor{black}{}

\textcolor{black}{A data set, chosen to easily recognize its middle
bit, is used to modulate a carrier at the central frequency of the
slow line. The GFSK modulator generates a synchronization signal.
The modulated signal alternatively propagates in the slow line and
in a microstrip line of the same overall length which is used as reference
measurement. After its transit in the slow line, or in the reference
line, the signal is demodulated using a Spectrum Analyzer (SPA) in
zero span mode with the same settings in both cases (reference level,
resolution filter, sweep time). This way we verified that the information
is correctly transmitted, and we measured the effective delay for
the information introduced by the slow line. The middle bit is identified
inside the data frame by demodulating the whole frame. Its start date
is measured as the middle of the rising time in both cases. This method
for measuring the information velocity is insensitive to any bias
that would result from some deformation of the signal. The velocities
measured this way can be found in Table~\ref{tab:Peformances} and
confirm the possibility to use the slow line to transmit data.}

\section{\textcolor{black}{Conclusion }}

\textcolor{black}{In this work, after having studied the possibilities
to delay a signal and verified that a low group velocity is a possible
method, we have demonstrated that a bandgap material permits to obtain
significant velocity reduction efficiently. A parametric study along
with theoretical considerations have allowed us to determine that
it is the velocity reduction bandwidth $\Delta f$ and the velocity
reduction factor that are the relevant parameters to design the line.
The theoretical study of a 1D hybrid bandgap metamaterial has shown
that it can be used to obtain the expected performances. Simulation
have allowed us to optimize the metamaterial line for real implementation.
A prototype has been realized using hybrid technology and presents
strong losses in spite of the use of PTFE woven glass substrate. We
have verified by simulation, after having measured the power dissipated
in the components, that the lost power is not radiated nor reflected.
The prototype presents the expected performances in terms of bandwidth
and group velocity in steady state conditions. In order to verify
that the variations of the transmission coefficient magnitude do not
prevent, by distorting the signal, to actually use the line to transmit
data, we have tested the data transmission capability. We thus verified
that the prototype allows data transmission at the expected low velocity.
Thus any delay could be obtained by increasing the number of cells.
Consequently realizing such delay line should present an effective
alternative to the use of optical fibers if realized using Monolitihic
Microwave Integrated Circuit (MMIC) technology.}

\begin{flushleft}
\textcolor{black}{\bibliographystyle{unsrt}}

\par\end{flushleft}
\end{document}